\documentclass{elsart}
\journal{EPSL}
\usepackage{amsmath}
\usepackage{amssymb}
\usepackage{graphicx}
\usepackage{epsfig}
\usepackage{marvosym}
\usepackage{wasysym}
\newcommand{\W}{\mbox{\scriptsize \Aquarius}}
\newcommand{\E}{\mbox{\scriptsize \Earth}}
\newcommand{\M}{\mbox{\scriptsize \Mars}}
\newcommand{\V}{\mbox{\scriptsize \Venus}}
\newcommand{\N}{\mbox{\hexstar}}

\begin{document}

\begin{frontmatter}

\title{A scaling law for aeolian dunes on Mars, Venus, Earth, and for subaqueous ripples}

\author{Philippe Claudin and Bruno Andreotti}

\address{Laboratoire de Physique et M\'ecanique des Milieux H\'et\'erog\`enes\\
UMR CNRS 7636\\
ESPCI, 10 rue Vauquelin, 75231 Paris Cedex 05, France.}

\date{\today}

\begin{abstract}
The linear stability analysis of the equations governing the evolution of a flat sand bed submitted to a turbulent shear flow predicts that the wavelength $\lambda$ at which the bed destabilises to form dunes should scale with the drag length $L_{\rm drag} = \frac{\rho_s}{\rho_f} \, d$. This scaling law is tested using existing and new measurements performed in water (subaqueous ripples), in air (aeolian dunes and fresh snow dunes), in a high pressure CO$_2$ wind tunnel reproducing conditions close to the Venus atmosphere and in the low pressure CO$_2$ martian atmosphere (martian dunes). A difficulty is to determine the diameter of saltating grains on Mars. A first estimate comes from photographs of aeolian ripples taken by the rovers Opportunity and Spirit, showing grains whose diameters are smaller than on Earth dunes. In addition we calculate the effect of cohesion and viscosity on the dynamic and static transport thresholds. It confirms that the small grains visualised by the rovers should be grains experiencing saltation. Finally, we show that, within error bars, the scaling of $\lambda$ with  $L_{\rm drag}$ holds over almost five decades. We conclude with a discussion on the time scales and velocities at which these bed instabilities develop and propagate on Mars.
\end{abstract}

\end{frontmatter}

Aeolian sand dunes form appealing and photogenic patterns whose shapes have been classified as a function of wind regime and sand supply \cite{FD79,PT90}. These dunes have martian `cousins' with very similar features \cite{nasa}. Grains can also be transported by water flows, and bed instabilities are commonly reported in flumes, channels and rivers \cite{Y92,CM96,CE01}. The same dynamical mechanisms control the formation of aeolian (both on Earth and Mars) dunes and subaqueous ripples\footnote{Subaqueous `ripples' are defined as patterns whose typical size is much smaller that the water height, whereas that of subaqueous `dunes' is comparable to depth.} from a flat sand bed. In a first section we describe the framework in which this instability can be understood. A unique length scale is involved in this description, namely the sand flux saturation length $L_{\rm sat}$, which scales on the drag length $L_{\rm drag} = \frac{\rho_s}{\rho_f} d$, where $d$ is the grain diameter and $\rho_s/\rho_f$ the grain to fluid density ratio. It governs the scaling of the wavelength $\lambda$ at which dunes or subaqueous ripples nucleate. 

However, attempting to test the scaling laws governing aeolian sand transport and dunes characteristic size, one faces a major problem: the grain size does not vary much from place to place at the surface of Earth and the grain to air density ratio is almost constant. This narrow range of grain size composing these dunes has a physical origin: large grains are too heavy to be transported by the wind and very small ones -- dust -- remain suspended in air. As shown by Hersen \emph{et al.} \cite{HDA02,H05}, small-scale barchan dunes can be produced and controlled under water, with a characteristic size divided by $800$ -- the water to air density ratio -- with respect to aeolian sand dunes. The martian exploration and in particular the discovery of aeolian patterns (ripples, mega-ripples and dunes) give the opportunity to get an additional point to check transport scaling laws, as the martian atmosphere is typically $60$ to $80$ times lighter than air. Finally, data from high pressure CO$_2$ wind tunnel reproducing conditions close to the Venus atmosphere \cite{GML84} as well as fresh snow barchan pictures \cite{MLWA98,Metal01} nicely complement and confirm the scaling relation.

A difficulty already risen in \cite{EC91} is to determine the diameter of saltating grains on Mars. A first estimate is derived from the analysis of photographs of the soil taken by the rovers Opportunity and Spirit next to ripple patterns, showing grains whose diameters are smaller than on Earth dunes \cite{Setal05}. We derive in the appendix the effect of cohesion and viscosity on the dynamic and static transport thresholds. It confirms that the small grains visualised by the rovers are experiencing saltation. Finally, we show that, within error bars, $\lambda$ scales with $L_{\rm drag}$ over almost five decades and discuss the corresponding time scales and velocities at which these bed instabilities develop and propagate.

\section{Dune instability}
\label{instab}

\begin{figure}[t!]
\begin{center}
\epsfig{file=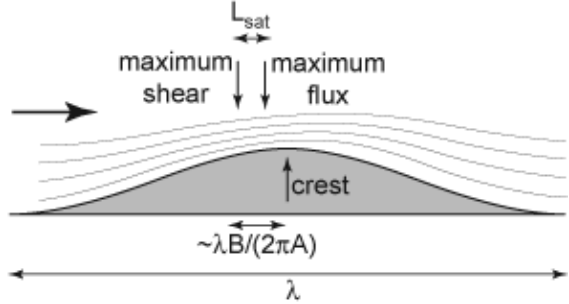}
\end{center}
\caption{Schematic of the instability mechanism showing the stream lines around a bump, the fluid flowing from left to right. A bump grows when the point at which the sand flux is maximum is shifted upwind the crest. The shift of the maximum shear stress scales on the size of the bump. The spatial lag between the shear and flux maxima is the saturation length $L_{\rm sat}$.
\label{InstabilityMechanism}}
\end{figure}

The instability results from the interaction between the sand bed profile, which modifies the fluid velocity field, and the flow that modifies in turn the sand bed as it transport grains. Let us consider a flat and symmetric bump as that shown in figure~\ref{InstabilityMechanism}. The fluid is accelerated on the upwind (stoss) side and decelerated on the downwind side. This is schematized by the flow lines in this figure: they are closer to each other above the bump. This results in an increase of the shear stress $\tau$ applied by the flow on the stoss side of the bump. Conversely, $\tau$ decreases on the lee side. Assuming that the maximum amount of sand that can be transported by a given flow -- the \emph{saturated} sand flux -- is an increasing function of $\tau$, erosion takes place on the stoss slope as the flux increases, and sand is deposited on the lee of the bump. If the velocity field was symmetric around the bump, the transition between erosion and deposition would be exactly at the crest, and this would lead to a pure propagation of the bump, without any change in amplitude (we call this the `$A$' effect, see below). In fact, due to the simultaneous effects of inertia and dissipation (viscous or turbulent), the velocity field is asymmetric (even on a symmetrical bump) and the position of the maximum shear stress is shifted upwind the crest of the bump (the `$B$' effect). In addition,  the sand transport reaches its saturated value with a spatial lag, characterized by the saturation length $L_{\rm sat}$. The maximum of the sand flux $q$ is thus shifted downwind  the point at which $\tau$ is maximum by a typical distance of the order of $L_{\rm sat}$.  The criterion of instability is then geometrically related to the position at which the flux is maximum with respect to the top of the bump: an up-shifted position leads to a deposition of grains before the crest, so that the bump grows.

The above qualitative arguments can be translated into a precise mathematical framework, see e.g. \cite{ACD02,ECA05}. For a small deformation of the bed profile $h(t,x)$, the excess of stress induced by a non-flat profile can be written in Fourier space as:
\begin{equation}
\hat{\tau} = \tau_0 (A + iB) k \hat{h},
\label{tau}
\end{equation}
where $\tau_0$ is the shear that would apply on a flat bed and $k$ is the wave vector associated to the spatial coordinate $x$. $A$ and $B$ are dimensionless functions of all parameters and of $k$ in particular. The $A$ part is in phase with the bed profile, whereas the $B$ one is out of phase, so that the modes of $h$ and $\tau$ of wavelength $\lambda$ have a spatial phase difference of the order of $\lambda B/ (2\pi A)$. This shift is typically of the order of $10\%$ of the length of the bump as $A$ and $B$ are typically of the same order of magnitude. Expressions for $A$ and $B$ have been derived by Jackson and Hunt \cite{JH75} in the case of a turbulent flow. As shown by Kroy \emph{et al.} \cite{KSH02}, $A$ and $B$ only weakly (logarithmically) depend on the wavelength so that they can be considered as constant for practical purpose. 

\begin{figure}[t!]
\begin{center}
\epsfig{file=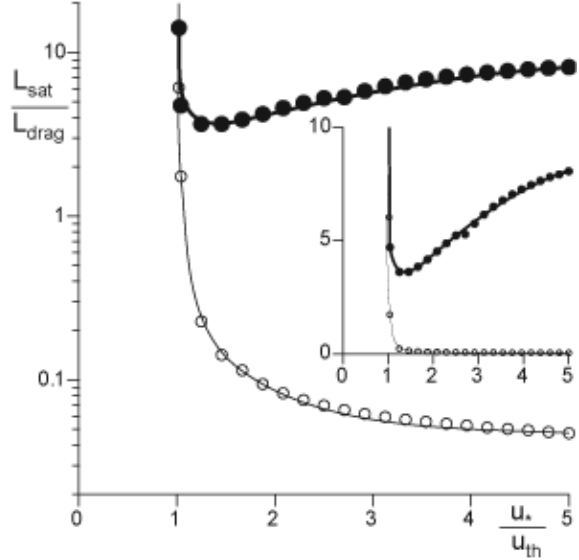}
\end{center}
\caption{Sand transport relaxation lengths in units of $L_{\rm drag}$ as a function of the rescaled wind velocity $u_*/u_{\rm th}$, as predicted by the theoretical model presented in \cite{A04}. The inset shows the same curves in linear scale. Starting from a vanishing flux, the sand transport first increases exponentially over a distance shown with $\circ$ symbols. The dominant mechanism during this phase is the ejection of new grains when saltons collide the sand bed. The spatial corresponding growth rate diverges at the threshold and rapidly decreases with larger $u_*$. As the number of grains transported increases, the wind is slowed down in the saltation curtain, until saturation. The distance over which the flux relaxes towards its saturated value is shown with {\Large $\bullet$} symbols. Except very close to the threshold, the dominant mechanism is the negative feedback of transport on the wind.
\label{Lsat}}
\end{figure}

If the shear stress is below the dynamical threshold $\tau_{\rm th}$, no transport is observed, hence the sand flux is null. Above this threshold, one observes on a flat bed a saturated flux $Q$ which is a function of $\tau_0$. The fact that a wind of a given strength can only transport a finite quantity of sand is due to the negative feedback of the grains transported by the fluid on the flow velocity profile -- the moving grains slow down the fluid. This saturation process of the sand flux is still a matter of research \cite{UH87,AH88,A04}. We refer the interested reader to \cite{A04,AH91} for a review discussion. For our present purpose, we need a first order but quantitative description of the saturated flux. Following \cite{A04}, Rasmussen \emph{et al.} wind tunnel data \cite{RIR96} are well described by the relationship:
\begin{equation}
Q = 25 \, \frac{\tau_0 - \tau_{\rm th}}{\rho_s} \, \sqrt{\frac{d}{g}}
\quad \mbox{if} \quad \tau_0 > \tau_{\rm th}
\quad \mbox{and} \quad Q=0 \quad \mbox{otherwise.}
\label{Qdetau}
\end{equation}
$g$ is gravitational acceleration. The prefactor $25$ has been adjusted to fit the data and is reasonably independent of the grain size $d$. The equivalent of the relation (\ref{tau}) for the saturated flux on a modulated surface $q_{\rm sat}$ can then be written as
\begin{equation}
\hat{q}_{\rm sat} = Q (\tilde{A} + i\tilde{B}) k \hat{h},
\label{qsat}
\end{equation}
where, by the use of relation (\ref{Qdetau}), the values of $\tilde{A}$ and $\tilde{B}$ simply verify $A/\tilde{A} = B/\tilde{B} = 1 - \tau_{\rm th}/\tau_0$.

As for any approach to an equilibrium state, there exists a relaxation length -- or equivalently a relaxation time -- scale associated with the sand flux saturation. This was already mentioned by Bagnold who measured the spatial lag needed by the flux to reach its saturated value on a flat sand patch \cite{B41}. A saturation length $L_{\rm sat}$ in dune models has been first introduced by Sauermann \emph{et al.} \cite{SKH01}, where the dependence of $L_{\rm sat}$ on $\tau$ and in particular its divergence as $\tau \to \tau_{\rm th}$ has been put phenomenologically in the description. In fact, the saturation length should \emph{a priori} depend on the mode of transport -- at least we are sure that it must exist in all the situations where there is an equilibrium transport. As there can be different mechanisms responsible for a lag before saturation (delay to accelerate the grain to the fluid velocity, delay due to the sedimentation of a transported grain \cite{C06}, delay due to the progressive ejection of grains during collisions, delay for the negative feedback of  the transport on the wind, delay for electrostatic effects between the transported grains and the soil, etc), the dynamics is dominated by the slowest mechanism so that $L_{\rm sat}$ is the largest among the different possible relaxation lengths. In the aeolian turbulent case, there exists a detailed theoretical analysis \cite{A04} providing the dependence of the saturation length on the shear velocity ({\Large $\bullet$} in figure~\ref{Lsat}). The curve roughly presents two zones. Very close to the threshold, the slowest process is the ejection of grains during collision. It is thus natural to have a divergence of the saturation length at the threshold \cite{SKH01} as the replacement capacity crosses $1$ (see the calculation of the dynamical threshold in Appendix). From just above the threshold -- say for $u_*/u_{\rm th} \gtrsim 1.05$ -- the saturation length gently increases (roughly linearly) with $u_*$. However, in the field, the mean wind strength varies from day to day, as seasons goes by. For practical purposes --~e.g. photograph analysis~-- it is thus of fundamental importance to define an effective saturation length, independent of the wind velocity. Fortunately, the velocity rarely exceeds $\sim 3~u_{\rm th}$, so that, in the range of interest, the velocity dependence on $L_{\rm sat}$ is subdominant. Our first important conclusion is thus that even though it should be remembered that $L_{\rm sat}$ slightly depends on $u_*/u_{\rm th}$ in lab experiments, the dominant parameters are the grain size $d$ and the sand to fluid density ratio. The theoretical analysis \cite{A04} shows that $L_{\rm sat}$ scales on the drag length $L_{\rm drag}= \frac{\rho_s}{\rho_f} \, d$, which is the length needed for a grain in saltation to be accelerated to the fluid velocity. It is worth noting that this inertial effect is however \emph{not} the mechanism limiting the saturation process (see above). Using the results of a field experiment performed in the Atlantic Sahara of Morocco \cite{ECA05}, the prefactor between $L_{\rm sat}$ and $L_{\rm drag}$ can be computed and gives:
\begin{equation}
L_{\rm sat} \simeq 4.4 \, \frac{\rho_s}{\rho_f} d.
\label{Lsat_turbu}
\end{equation}
This result is also consistent with Bagnold's data \cite{B41}. Note that the saturation length has never been measured directly in other situations  (neither under water nor in high/low pressure wind tunnels). 

The linear stability analysis of the coupled differential equations of this framework has been performed in \cite{ACD02}. In particular, the wavenumber corresponding to the maximum growth rate is given by
\begin{equation}
k_{\rm max} \, L_{\rm sat} = X^{-1/3}-\frac{X^{1/3}}{3} \;\;\; {\rm with}\;\;\; X=\frac{3\tilde{B}}{\tilde{A}}\;\left[-3 + \sqrt{3\left(1+ (\tilde{A}/\tilde{B})^2\right)}\right].
\label{kmax}
\end{equation}
For typical values of $\tilde{A}$ and $\tilde{B}$ determined in the barchan dune context \cite{ECA05}, we have $\lambda_{\rm max} = 2\pi/k_{\rm max} \sim 12 L_{\rm sat}$. Note that we have $A/B=\tilde{A}/\tilde{B}$, so that $k_{\rm max} \, L_{\rm sat}$ is independent of $\tau_0$ and $\tau_{\rm th}$. Using measurements in water (subaqueous ripples), in air (aeolian dunes) and in the CO$_2$ atmosphere of Mars and of the Venus wind tunnel, we now investigate the scaling relation between $\lambda_{\rm max}$ and $L_{\rm drag}$. To do so, we need to first solve the controversy concerning the size of saltating grains on Mars.
\begin{figure}[t!]
\begin{center}
\epsfig{file=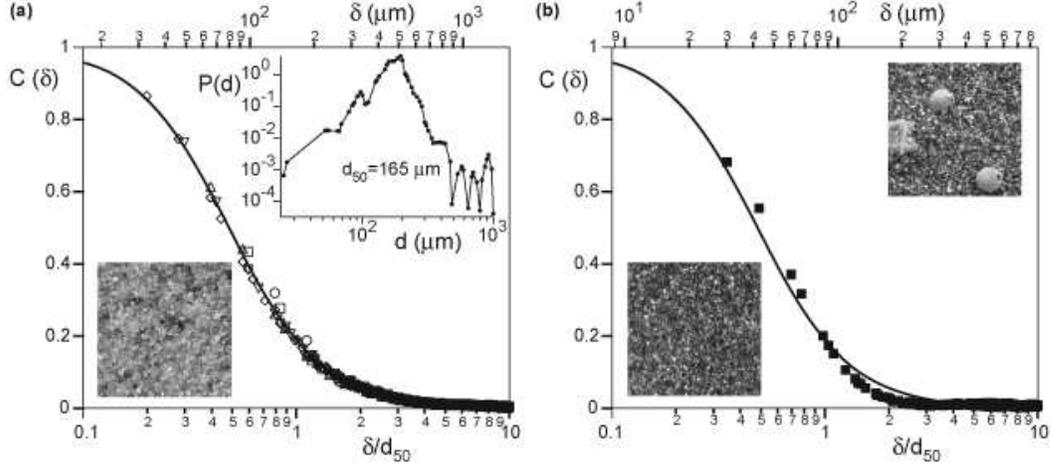,width=\linewidth}
\end{center}
\caption{Measurement of the typical grain size from the auto-correlation function $C(\delta)$ of a photograph of the granular bed. {\bf a} Sample of aeolian sand from the mega-barchan of Sidi-Aghfinir (Atlantic Sahara, Morocco), of typical diameter $d \simeq 165~\mu$m. Top inset: Probability distribution function of the grain diameter $d$, weighted in mass. The narrow distribution around the maximum of probability is characteristic of the aeolian sieving process. Bottom inset: $200 \times 200$~pix$^2$ zoom on a typical photograph on which the computation of the autocorrelation function has been performed. The image resolution is here $5$~pix/$d$. {\bf b} Sample of sand from Mars (at rover Spirit's landing site: $16.6^\circ$ S, $184.5^\circ$ W). Top inset: photograph showing the presence of milimetric grains as well as much smaller ones which are expected to be the saltons. Bottom inset: same as {\bf a}, but the typical image resolution is $3$~pix/$d$.
\label{Granulo}}
\end{figure}

\section{Size of saltating grains on Mars}

The determination of the typical size of the grains participating to saltation on martian dunes is a challenging issue. First, the dunes seem to evolve very slowly or may even have become completely static. Second, no sample of the matter composing the bulk of the dunes is available. Third, we do know from the observation of dunes on Earth that they can be covered by larger grains that do not participate to the transport in saltation. With the two rovers Opportunity and Spirit, we now have direct visualisations of the soil \cite{Setal05}, and in particular of clear aeolian structures like ripples\footnote{Contrarily to dunes, aeolian sand ripples form by a screening instability related to geometrical effects \cite{B41,A90}.} or nabkhas\footnote{As there is a reduction of pressure in the lee of any obstacle, sand accumulates in the form of streaks aligned in the direction opposite to the wind (shadow dunes). These structures are called nabkhas \cite{CWG93}.}. Unfortunately, these structures did not lie on the surface of a dune. We thus analyze the available photographs, assuming that, like on Earth, the size of the grains participating in saltation does not vary much from place to place.

\subsection{Direct measurement of grain sizes}

The photographs freely accessible online are not of sufficiently good resolution to determine the shape and the size of each grain composing the surface. Besides, one has to be careful when analyzing such pictures as part of what is visible at the surface corresponds to grains just below it and partly hidden by their neighbours. We have thus specifically developed a method to determine the average grain size in zones where the grains are reasonably monodispersed, when the resolution is typically larger than $3$ pixels per grain diameter. This measure can be deduced from the computation of the auto-correlation function $C(\delta)$ of the picture, which decreases typically over one grain size. More precisely, we proceed with the following procedure:\\
$\bullet$ The zones covered by anything but sand (e.g. gravels or isolated larger grains) are localized and excluded from the analysis.\\
$\bullet$ Because in natural conditions the light is generally inhomogeneous, we perform a local smoothing of the picture with a gaussian kernel of radius $\simeq 10~d$. The resulting picture is subtracted from the initial one.\\
$\bullet$ We compute the local standard deviation of this image difference with the same gaussian kernel and produce, after normalisation, a third picture $\mathcal{I}$ of null local average and of standard deviation unity.\\
$\bullet$ The auto-correlation function $C(\delta)$ is computed on this resulting picture, averaging over all directions:
\begin{equation}
\label{Ddedelta}
C(\delta) = \frac{\sum_{k,l/k^2+l^2=\delta^2} \sum_{i,j} \mathcal{I}_{i,j} \mathcal{I}_{i+k,j+l}}
                            {\sum_{k,l/k^2+l^2=\delta^2} \sum_{i,j} 1}
\end{equation}

Figure \ref{Granulo}a shows the curve $C(\delta)$ obtained from a series of photographs of aeolian sand sampled on a terrestrial  dune. For all the resolutions used (between $1$ and $5$~pix/$d$), the data collapse on a single curve, which is thus characteristic of the sand sample. The top inset shows the distribution of size, weighted in mass, in log-log representation. It presents a narrow peak around the $d_{50}$ value. The autocorrelation function decreases from $1$ to $0$ over a size of the order of $d_{50}$ (Lorenzian fit). This is basically due to the fact that the color or the gray level at two points are only correlated if they are inside the same grain. It is thus reasonable to assume that the autocorrelation curve is a function of $\delta/d_{50}$ only. As a matter of fact, photographs of two samples of comparable polydispersity are similar once rescaled by the mean grain diameter. We will thus use the curve obtained with aeolian sand sampled on Earth as a reference to determine the size of Martian grains. Figure~\ref{Granulo}b shows the autocorrelation curve obtained from the colour image taken by the rover Spirit at its landing site. $C(\delta)$ decreases faster than the reference curve but by tuning the value of the martian $d_{50}$, one can superimpose the Mars data with the solid line fairly well. From this picture as well as a series of gray level photographs taken by the rover Opportunity, we have estimated the diameter of the grains composing these aeolian formations to $87 \pm 25~\mu$m.
\begin{figure}[t!]
\begin{center}
\epsfig{file=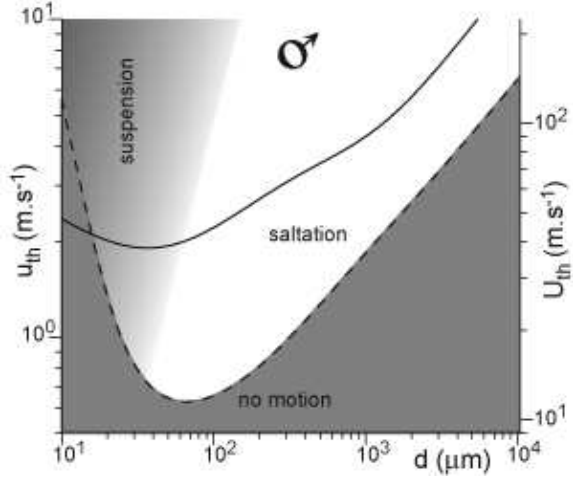}
\end{center}
\caption{Diagram showing the mode of transport on Mars as a function of the grain diameter $d$ and of the turbulent shear velocity $u_{\rm th}$ (left) or of the wind speed $U_{\rm th}$ at $2$~m above the soil (right). Below the dynamical threshold (dashed line), no grain motion is observed. A grain at rest on the surface of the bed starts moving, dragged by the wind, when the velocity is above the static threshold (solid line). Between the dynamical and static thresholds, there is a zone of hysteresis where transport can sustain due to collision induced ejections. The progressive transition from saltation to suspension as wind fluctuations become more and more important is indicated by the gradient from white to gray color. See appendix for more details on the derivation of this graph.
\label{ShieldsMars}}
\end{figure}

The first measurement of martian grain sizes dates back to the Viking mission in the 70s. On the basis of thermal diffusion coefficient measurements, Edgett and Christensen have estimated the grain diameter to be around $500~\mu$m and at least larger than those composing dunes on Earth \cite{EC91}. This size corresponds to larger grains than saltons such as those shown in the top inset of figure~\ref{Granulo}b. In agreement with our findings, wind tunnel experiments performed in `martian conditions' \cite{GLWIP80} have shown that grains around $100~\mu$m are the easiest to dislodge from the bed. We shall come back to this point later on.

Several theoretical investigations of the size of aeolian martian grains have been conducted, starting with the work of Sagan and Bagnold \cite{SB75}. In that paper the authors argue that, as Mars is a very arid planet, the cohesion forces due to humidity can be neglected and proposed a cohesion-free computation which predicts that very small grains (typically of one micron) may be put into saltation. Miller and Komar \cite{MK77} also followed this cohesion-free approach and proposed static threshold curves with no turnup on small particle size. It was soon realized that cohesion forces can occur for reasons other than humidity --~namely van der Waals forces~-- and several authors proposed (static) transport threshold curves with a peaked minimum around $100~\mu$m \cite{IPGW76,PHGI76,IW82,SL00,CGH04}. However in these papers, cohesion is treated in an empirical fashion, with the assumption that van der Waals forces lead to an attractive force proportional to the grain diameter with a prefactor independent of $d$.

We have recomputed the Martian transport diagram (figure~\ref{ShieldsMars}) using a new derivation of the transport thresholds. It takes into account the hysteresis between the static and dynamic thresholds, the effect of viscosity and --~in a more rigorous way~-- the effect of cohesion. As this derivation, although consequent and original, is not the central purpose of the present paper which is devoted to the test of the dune scaling law, we have developed and discussed it in appendix.

We wish to solely discuss here the figure~\ref{ShieldsMars}, which is useful to prove that the $87~\mu$m sized grains can be transported in saltation. The first striking feature of the Martian transport diagram is the huge hysteresis between the dynamic and static thresholds. Compared to aeolian transport on Earth (figure~\ref{ShieldsWaterAir} in the Appendix) for which the static threshold is typically $50\%$ above the dynamic threshold, they are separated on Mars by a factor larger than $3$. Quantitatively, if the static threshold is very high compared to the typical wind velocities on Mars ($\sim 150~$km/h at $2~$m above the soil), the dynamical threshold is  only of the order of $\sim 45~$km/h at $2~$m. From images by the Mars Orbiter Camera (MOC) of reversing dust streaks, Jerolmack \emph{et al.} \cite{JMGFW06} have estimated that modern surface winds can reach velocities as large as $150$~km/h. Looking at figure~\ref{ShieldsMars}, it can be seen that at such velocities, most of the grain below $100~\mu$m can be suspended and that even millimetric grains can be entrained into saltation. Even in less stormy conditions, sand transport should not be as unfrequent as one could expect, even though the large amplitude of the hysteresis implies an intermittency of sand transport. This suggests that Martian dunes are definitively active.

\section{A dune wavelength scaling law}

Keeping in mind that the wavelength $\lambda$ that spontaneously appears when a flat sand bed is destabilized by a turbulent flow scales on the flux saturation length, the aim of the paper is to plot $\lambda$ against $L_{\rm drag}$ in the different situations mentioned in the introduction: aqueous ripples, waves on aeolian dunes, both on Earth and Mars, fresh snow dunes and microdunes in the Venus wind tunnel -- numerical values of the parameters corresponding to these different situations are summarized in table \ref{tab_compare}.

\begin{table}[t!]
\begin{center}
\begin{tabular}{|c|c|c|c|c|c|}
\hline
 & Earth \E & Mars \M & water \W & snow \N & `Venus' \V \\
\hline
$g$ (m/s$^2$) & $9.8$& $3.7$  & $9.8$  & $9.8$& $9.8$\\
$\lambda$ & $20$~m & $600$~m  & $2$~cm & $15$--$25$~m & $10$--$20$~cm\\
$d$~($\mu$m)& $165$ -- $185$ & $87$& $150$ & $1500$ & $110$\\
$\rho_f$~(kg/m$^3$)& $1.2$ & $1.5$ -- $2.2~10^{-2}$ & $10^{3}$&$1.2$&$61$\\
$\rho_s$~(kg/m$^3$)& $2650$ & $3000$ & $2650$ & $360$ & $2650$\\
$\nu$~(m$^2$/s)&$1.5~10^{-5}$& $6.3~10^{-4}$ &$10^{-6}$&$1.5~10^{-5}$&$2.5~10^{-7}$\\
\hline
\end{tabular}
\vspace*{0.3cm}
\end{center}
\caption{Comparison of different quantities (gravity $g$, initial wavelength of bed instability $\lambda$, diameter of saltons $d$ as well as fluid and sediment densities $\rho_f$ and $\rho_s$) in the air (Earth), in the martian and Venus wind tunnel CO$_2$ atmospheres and in water. As the temperature at the surface of Mars can vary by an amplitude of typically $100$~K between warm days and cold nights, the density of the atmosphere displays some variation range.
\label{tab_compare}}
\end{table}
\begin{figure}[p]
\begin{center}
\epsfig{file=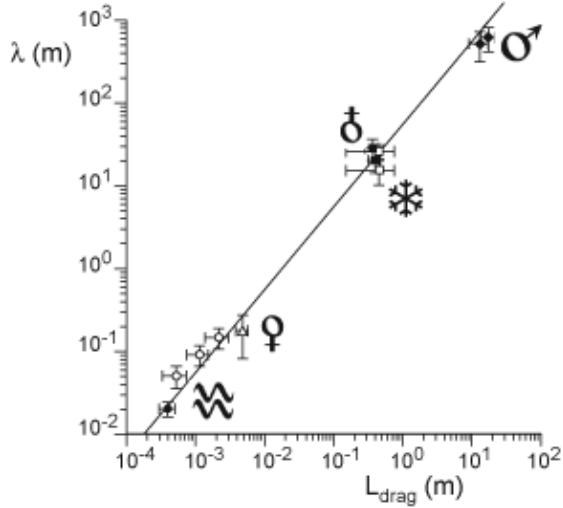}
\end{center}
\caption{Average wavelength as a function of the drag length. The two black squares (resp. diamonds) labeled `Earth' ($\E$) (resp. `Mars' ($\M$)) come from the two different histograms of figure \ref{histo}a (resp. \ref{histo}b). The three white circles are the under water ($\W$) Coleman and Melville's data \cite{CM96,CE01}, whereas the black circle is that from Hersen \emph{et al.}'s experiments \cite{HDA02,H05}. The white triangle has been computed from the Venus ($\V$) wind tunnel study \cite{GML84}. Snow ($\N$) dune photos (see figure \ref{barchans_snow}) have been calibrated and complemented with data from \cite{MLWA98,Metal01,Hpc}, and give the white squares.
\label{Dunes_scalingMTW}}
\end{figure} 
\begin{figure}[p]
\begin{center}
\epsfig{file=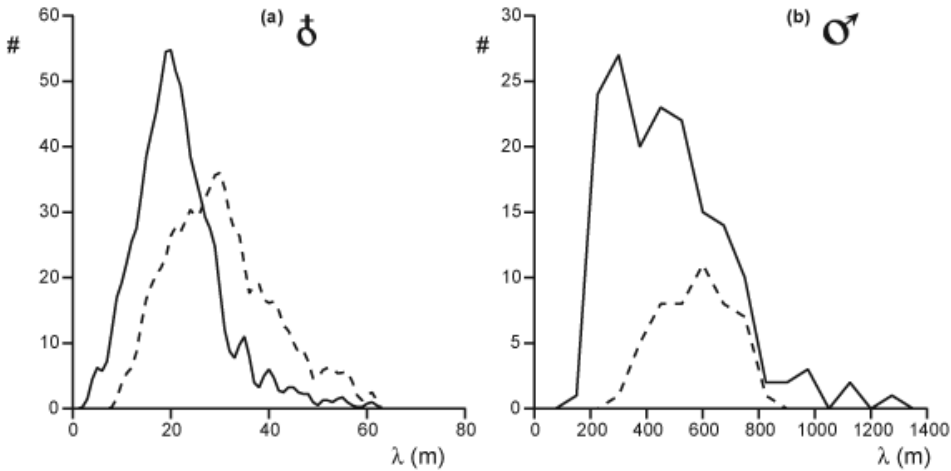,width=\linewidth}
\end{center}
\caption{Histograms of the wavelength measured in the Atlantic Sahara (Morocco) and in Mars southern hemisphere (mainly, but not only, in the region $320$--$350^\circ$W, $45$--$55^\circ$S). {\bf (a)} Earth ($\E$). Solid line: wavelengths systematically measured on barchan dunes located in a 20 km $\times$ 8 km zone; Dash line: wavelengths measured on the windward side of a mega-barchan whose surface is permanently corrugated and where some coarsening occurs. {\bf (b)} Mars ($\M$). Solid line: wavelengths measured on dunes in several craters (e.g. Rabe, Russell, Kaiser, Proctor, Hellespontus); Dash line: histogram restricted to Kaiser crater ($341^\circ$W, $47^\circ$S). Averaged values of $\lambda$ are $20$~m (solid line on panel a), $28$~m (dash line on panel a), $510$~m (solid line on panel b) and $606$~m (dash line on panel b).
\label{histo}}
\end{figure}
\begin{figure}[p]
\begin{center}
\epsfig{file=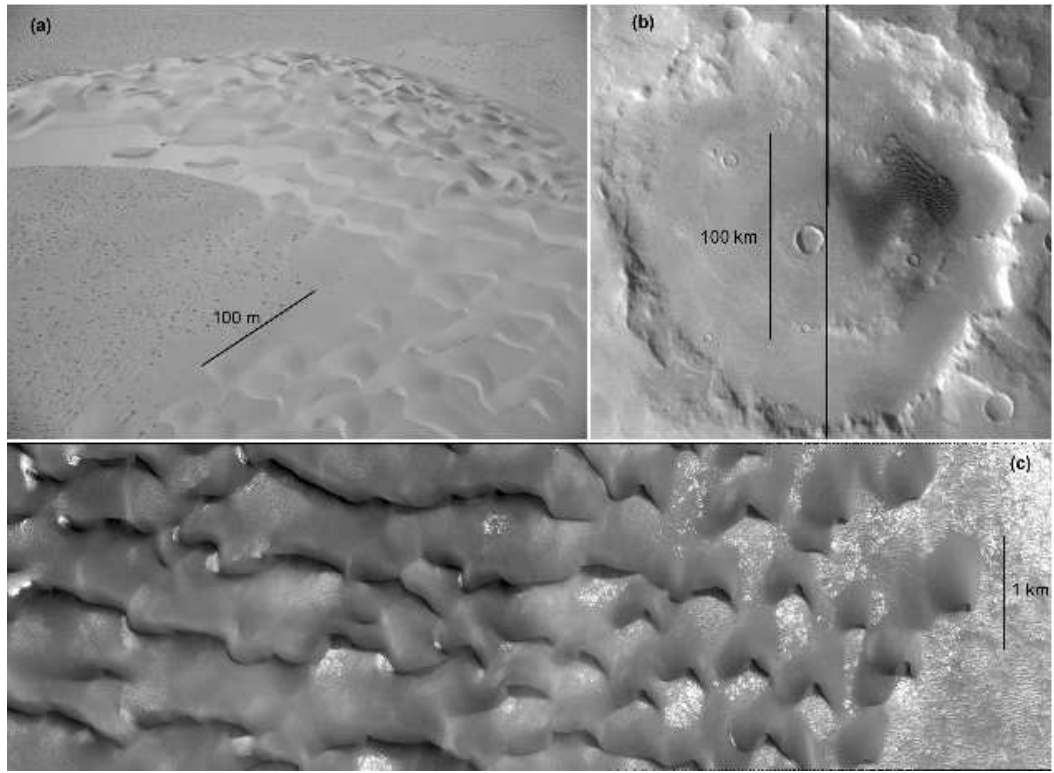,width=\linewidth}
\end{center}
\caption{Comparison of dune morphologies on Earth and on Mars. {\bf a} Mega-barchan in Atlantic Sahara, Morroco. {\bf b} `Kaiser' crater on Mars ($341^\circ$W,$47^\circ$S). {\bf c} Closer view of the Kaiser crater dunes. As on Earth, small barchans are visible on the side of the field.
\label{ComparMarsTerre}}
\end{figure}
\begin{figure}[p]
\begin{center}
\epsfig{file=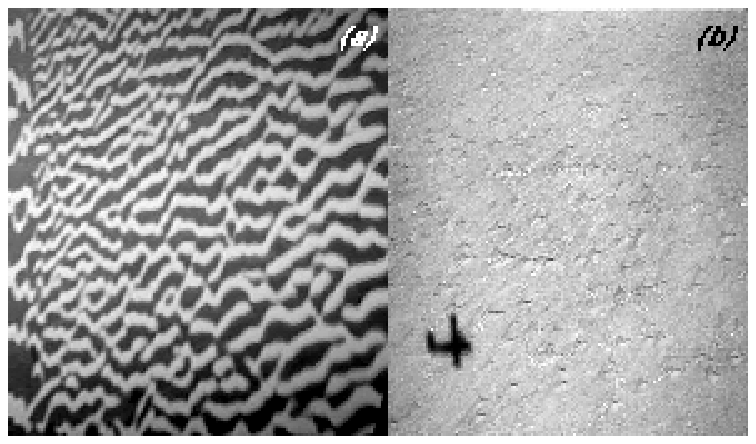}
\end{center}
\caption{Fresh snow aeolian dunes on ice. {\bf (a)} Transverse dunes on the iced baltic sea (credits Bertil H\aa kansson, Swedish Meteorological and Hydrographical Institute/Baltic Air-Sea-Ice Study). The wavelength is around $15$~m \cite{Hpc}. {\bf (b)} Snow barchan field in Antartica (credits Stephen Price, Department of Earth and Space Sciences, University of Washington). The shadow of the Twin Otters, which has has a wing span of $19.8$~m and a length of $15.7$~m, gives the scale. In both pictures, the perspective has been corrected to produce a pseudo-aerial view.
\label{barchans_snow}}
\end{figure}

Our reference data point in figure \ref{Dunes_scalingMTW} (aeolian dunes on Earth) has been obtained after an extensive work on barchan dunes in the Atlantic Sahara of Morocco. In a recent paper~\cite{ECA05}, we have shown that perturbations such as wind changes generate waves at the surface of dunes by the linear instability described above. Direct measurements of the wavelength are reported in Figure~\ref{histo}a in an histogram. They give $\lambda \sim 20$~m for waves on the flanks of medium sized dunes (solid line) and $\sim 28$~m on the windward side of a mega-barchan (dashed line), where some pattern coarsening occurs. $L_{\rm sat} \sim 1.7$~m was measured independently in the same dune field, and these aeolian grains of $180~\mu$m lead to a drag length of $40$~cm. 

The study of small scale barchans under water \cite{HDA02,H05} provides the second data point: measurements give $\lambda \sim 2$~cm with glass beads of size $d=150~\mu$m, a size which leads to a drag length of $400~\mu$m. This point is the only one in the literature for which it is specified that: (i) the wavelength has been measured during the linear stage of the instability; (ii) the sand bed destabilizes in a homogeneous way and not starting at the entrance of the set-up, due to strong disturbances there; (iii) the height of the tank is much larger than the wavelength. We have added three other underwater points in figure~\ref{Dunes_scalingMTW}, from Coleman and Melville's data \cite{CM96,CE01}.The slight discrepancy with the straight line (these points are all above the line) can be explained by the three points risen above: it may be induced from \cite{CM96} that (i) the initial stage is not clearly resolved; (ii) the waves seem to nucleate on defects as if the transition was sub-critical; (iii) the flume is only $0.28$~m deep, a height which is comparable to the observed wavelengths. 

Recent photos of martian dunes \cite{nasa} such as that in figure~\ref{ComparMarsTerre} lead to an estimate of the value of $\lambda \sim 600$~m on Mars. We focused on the dunes found in craters of the southern hemisphere. For comparison, we have measured wavelengths on dunes in several craters (e.g. Rabe, Russell, Kaiser, Proctor, Hellespontus) and also produced a histogram restricted to Kaiser crater, see figure~\ref{histo}b. As for the drag length, we used the value for the grain diameter estimated in the previous section. The density of martian grains is similar to or slightly higher \cite{JMGFW06} than that of terrestrial grains. The value of the density of the martian atmosphere however varies within some range as the temperature on Mars surface can change by an amplitude of typically $100$~K between warm days and cold nights. In the end, it gives a value of $L_{\rm drag}$ between $13$ and $17$~m.

The fourth data point is obtained from Greeeley et al.'s experiment in a high pressure CO2 wind tunnel \cite{GML84}, which gives for $L_{\rm drag}$ a value we expect on Venus. The observed wavelength first decreases from $18~$cm to $8~$cm with the wind speed from $u_*=u_{\rm th}$ to $u_*=1.6~u_{\rm th}$ and then increases up to $27~$cm at  $u_*=2.1~u_{\rm th}$. Above this value, the flat sand bed was again found to be stable. Although some of these features are reminiscent of the $L_{\rm sat}$ curve discussed above (figure~\ref{Lsat}), this series of experiments is also questionable: nothing is reported about the nature of the destabilization (homogeneous appearance of the pattern or not) nor on the maturity of the pattern when the wavelength was measured (coarsening?). Remarkably, the crude averaged of the provided data lies on the master curve, even though the smallest wavelength is less than half the predicted value.  

As for the Venus wind tunnel, the fifth data points are not very precise and correspond to fresh snow dunes formed on Antartic sea ice and on Baltic sea ice. The typical wavelength $\lambda$ of transverse dunes (for instance in figure~\ref{barchans_snow}{\bf (a)} ranges from $15$~m to $25$~m \cite{Hpc}. Whenever barchan dunes form, they are typically $5$~m to $10$~m long and are separated by  $\sim 20$~m (see figure~\ref{barchans_snow}{\bf (b)}. This is very similar with aeolian ones, although snow barchan dunes look more elongated.  The determination of $L_{\rm drag}$ is more problematic as the snow density rapidly increases once fallen. Snow dunes are rather rare and probably form only at the surface of ice by strong wind, when the snow can remain fresh and not very cohesive. To get the numbers, we have used the measurements performed close to the Antartic dunes by Massom \emph{et al.} \cite{MLWA98,Metal01}. Although one may find somehow disappointing that on the graph of figure~\ref{Dunes_scalingMTW} these data points are located very close to the aeolian sand dunes, they are particularly interesting as they show that one can keep the same $\lambda$ by changing simultaneously $\rho_s$ and $d$ in a compensating way.

Globally, we obtain a consistent scaling law $\lambda \simeq 53L_{\rm drag}$ that covers almost five decades (figure~\ref{Dunes_scalingMTW}). We wish to emphasize again that we do not claim to capture all the dependences on this plot, but that $L_{\rm drag}$ is the dominant scaling factor. As shown in figure~\ref{Lsat}, we expect subdominant dependences on the wind speed, on finite size effects, etc... Note that this analysis explains for example why martian dunes whose sizes are of the order of a kilometer are not `complex' or `compound' as their equivalents on Earth \cite{BGM79}. Following our scaling relation, they correspond to the small terrestrial dunes and the whole dune field on the floor of a crater should rather be considered as a complex martian dune. A `similarity law' for the size of (developed) dunes with $L_{\rm sat}$ was schematically drawn by Kroy \emph{et al.} \cite{KFO05}, supported by the existence of centimetric barchans under water \cite{HDA02}. However, this similarity was announced to fail with Martian dunes. In fact, the scaling of large dunes (i.e. whose sizes are much larger than the elementary length $\lambda$) with $L_{\rm drag}$ (or $L_{\rm sat}$) is far from obvious as the size selection of barchans involves secondary instabilities related to collisions and fluctuations of wind direction~\cite{ECA05,HEAACD04,EAC06}.

Finally, we would like to end this section with the prediction of  the wavelength at which a flat sand bed should destabilize on Titan where (longitudinal or seif) dunes have been recently discovered \cite{Letal06}. The atmosphere there is approximately four times denser than on Earth, and the grains are believed to be made of water ice \cite{LLGF95}. Computing the threshold curves (not shown), one observes that the dynamical and static thresholds are almost identical and that the minimum of threshold shear stress is reached for $160~\mu$m. Using this size for the grain diameter a density ratio of the order of $200$, we find a centimetric drag length. Following the scaling law of figure \ref{Dunes_scalingMTW}, this would lead to a dune wavelength between $1$ and $2$ meters. Note that, unfortunately, this is well below the resolution of Cassini radar.

\section{Discussion: time scales and velocity for the martian dunes}

Now that we have this unique length scale at hand for dunes, it is interesting to address the question of the corresponding growth time scales and propagation velocities. The linear stability analysis \cite{ACD02,ECA05} tells that the growth rate $\sigma$ and the propagation velocity $c$ of bedforms are related to the wavenumber $k=2\pi/\lambda$ as
\begin{equation}
\sigma(k)  =  Q k^2 \, \frac{\tilde{B}-\tilde{A} kL_{\rm sat}}{1+(kL_{\rm sat})^2} 
\qquad \mbox{and} \qquad
c(k)  =  Q k \, \frac{\tilde{A}+\tilde{B}kL_{\rm sat}}{1+(kL_{\rm sat})^2},
\label{sigmaetcdek}
\end{equation}
where $Q$ is the saturated sand flux --~assumed constant~-- over a flat bed. In order to make these relations quantitative for the bedforms on Earth and Mars, we need an effective time averaged flux $\bar{Q}$. Recall from equation (\ref{Qdetau}) that $Q$ can be related to the shear velocity $u_*$. For simplicity, we suppose that there is sand transport ($u_* > u_{\rm th}$) a fraction $\eta$ of the time and that the shear stress is then constant and equal to $(1+\alpha) \rho_f u_{\rm th}^2$. The time averaged sand flux can thus be effectively expressed as
\begin{equation}
\bar{Q} = 25\, \alpha \, \eta \, \sqrt{\frac{d}{g}} \, \frac{\rho_f}{\rho_s} u_{\rm th}^2 .
\end{equation}
The values of the coefficients $\tilde{A}$ and $\tilde{B}$ which come into equation (\ref{sigmaetcdek})  depend on the excess of shear above the threshold as $\tilde{A}/A = \tilde{B}/B = (1+\alpha)/\alpha$. The timescale over which an instability develops is that of the most unstable mode. This means that $\sigma$ and $c$ should be evaluated at $k=k_{\rm max}$ (see equation (\ref{kmax})) and thus scale as:
\begin{eqnarray}
\label{sigmadek2}
\sigma & \propto & \frac{\bar{Q}}{L_{\rm sat}^2} \propto
(1+\alpha) \, \eta \, \left(\frac{\rho_f}{\rho_s} \right)^3 \, \frac{u_{\rm th}^2}{g^{1/2} d^{3/2}},\\
\label{cdek2}
c & \propto & \frac{\bar{Q}}{L_{\rm sat}} \propto (1+\alpha) \, \eta \, \left(\frac{\rho_f}{\rho_s} \right)^2
\frac{u_{\rm th}^2}{g^{1/2} d^{1/2}} \,  .
\end{eqnarray}
Using meteorological data and measurements of the average dune velocities, we estimated that $Q_{\E}$ is between $60$ to $90$~m$^2$/year \cite{ECA05} and  $\eta_{\E}$ between $65$\% and $85$\%. This gives an effective value $\alpha_{\E}$ between $1.5$ and $2$. Calculating explicitly the prefactors, we get  a growth time $\sigma^{-1}_{\E} \sim 2$~weeks and $c_{\E} \sim 200$~m/year, values which are consistent with direct observation. Note that time scales for individual fully developed dunes depend on their size, but are also proportional to $1/\bar{Q}$ \cite{HEAACD04}.

In order to compare these terrestrial values to those on Mars, we need to estimate the different factors which come into the expressions (\ref{sigmadek2}) and (\ref{cdek2}). Recall that, on Earth, for grains of $165~\mu$m, we have $u_{\E {\rm th}} \sim 0.2~$m/s. The corresponding value for the $87~\mu$m sized Martian grains can be found on figure~\ref{ShieldsMars} and reads $u_{\M {\rm th}} \sim 0.64~$m/s. The Mars to Earth ratios for the fluid and grain densities, as well as the gravity acceleration are known. As discussed in the calculation of the transport thresholds (see Appendix) one expects that $\alpha_{\M}$ is of order unity and for simplicity we simply take $\alpha_{\M} = \alpha_{\E}$. Finally, the most speculative part naturally concerns the value of $\eta_{\M}$. Assuming that the winds on Mars are similar to those on Earth, we computed on our wind data from Atlantic Sahara the fraction of time during which $u_* > 0.64~$m/s and got $\sim 3\%$. This value corresponds to few days per year and is probably realistic as, besides, the soil of Mars is frozen during the winter season. With these numerical values, we find that $\sigma_{\M}$ is smaller than $\sigma_{\E}$ by more than five decades. In other words, the typical time over which we could see a significant evolution of the bedforms on martian dunes is of the order of tens to hundreds centuries. Similarly, the ratio $c_{\M}/c_{\E}$ is of the order of $10^{-4}$. Note that these Mars to Earth time and velocity ratios are proportional to $\eta_{\M}$, so that larger or smaller values for this speculative parameter do not change dramatically the conclusions, the main contribution being the density ratio to the power $2$ or $3$ (see equations (\ref{sigmadek2}--\ref{cdek2})). Therefore, as some satellite high resolution pictures definitively show some evidence of aeolian activity -- e.g. avalanche outlines -- it may well be that the martian dunes are fully active but not significantly at the the human scale.

\section{Conclusion}

As for the last section, we would like to conclude the paper with a summary of the status of the numerous hypothesis and facts we have mixed and discussed. Although coming from recent theoretical works, it is now well accepted that the wavelength $\lambda$ at which dunes form from a flat sand bed is governed by the so-called saturation length $L_{\rm sat}$. However, the dependences of $L_{\rm sat}$ with the numerous control parameters (Shields number, grain and flow Reynolds numbers, Galileo number, grain to fluid density ratio, finite size effects, etc) is still a matter of debate. In this paper, we have collected measurements of $\lambda$ in various situations that where previously thought of as disconnected: subaqueous ripples, micro-dunes in Venus wind tunnel, fresh snow dunes, aeolian dunes on Earth and dunes on Mars. We show that the averaged wavelength (and thus $L_{\rm sat}$) is proportional to the grain size times the grain to fluid density ratio. This does not preclude sub-dominant dependencies of $L_{\rm sat}$ with the other dimensionless parameters. In  each situation listed above, we have faced specific difficulties.\\
{\it Subaqueous ripples~---~}The transport mechanism under water, namely the direct entrainment by the fluid, is different from the four other situations. The saturation process could thus be very different in this case. The longest relaxation length could for instance be  controlled by the grain sedimentation, as recently suggested by Charru~\cite{C06}. As there is also negative feedback of transport on the flow, the destabilization wavelength would be larger than our prediction. The formation of subaqueous ripples has remained controversial mostly because the experimental data are very dispersed. Most of them suffer from finite size effects (water depth or friction on lateral walls), from uncontrolled entrance conditions leading to an inhomogeneous destabilization, or from a late determination of the wavelength after a pattern coarsening period. For the furthest left data point only, we are certain that all these problems were avoided.\\
{\it `Venus' micro-dunes~---~}This beautiful experiment gives the most dispersed data of the plot. Part of it may be due to  the defects listed just above. The complex variation of the wavelength with the wind speed observed experimentally could be interpreted as a sub-dominant dependance of the saturation length on the Shields number. However, not only the size of the dunes changes but also their shapes. This is definitely the signature of a second length scale at work.\\
{\it Fresh snow dunes~---~}On the basis of existing photographs, we have clearly identified snow dunes pattern resembling sand aeolian destabilization ones. They form under strong wind, on icy substrate. The obvious difficulty is that the precise state of the snow flakes during the dune formation involves complex thermodynamical processes that do not exist in the other cases.\\
{\it Aeolian dunes on Earth~---~}Although the instability takes place in the field under varying wind conditions, the resulting wavelength is very robust from place to place and time to time. We consider this point as the reference one in the plot.\\
{\it Martian dunes~---~}The specific difficulty of the Martian case does not come from the wavelength measurements, as the available photographs are well resolved in space, but rather from the prior determination of the diameter $d$ of the grains involved. The controversy comes from early determinations based on thermal diffusivity measurements, concluding that dunes are covered by large ($500~\mu$m) grains. Such large grains would lead to a significant shift of the data point towards the right of figure~\ref{Dunes_scalingMTW}. A large part of this paper is consequently devoted to an independent determination of $d$, based on the analysis of the Martian rovers photographs. Our determination is very different: $87\pm25~\mu$m. We have computed the sand transport phase diagram with more subtleties than in the available literature (including in particular hysteresis and cohesion) and shown that this size corresponds to saltating grains. A directly related controversial point  was the state of the Martian dunes (still active or fossile). We have shown that moderately large winds can transport the grains (the dynamical threshold is much lower than the static one) but that the characteristic time scale over which dunes form is five orders of magnitude larger than on Earth.

\rule[0.1cm]{3cm}{1pt}\\
The Moroccan wavelength histogram has been obtained in collaboration with Hicham Elbelrhiti. We thank \'Eric Cl\'ement and \'Evelyne Kolb for useful advices on the way cohesion forces can be estimated. We thank Fran\c cois Charru and Douglas Jerolmack for discussions. We thank Brad Murray for a careful reading of the manuscript. Part of this work is based on lectures given during the granular session of the Institut Henri Poincar\'e in 2005. This study was financially supported by an `ACI Jeunes Chercheurs' of the french ministry of research.

\newpage
\appendix

\section{A model for transport thresholds, including cohesion}
We have directly measured the grain diameter ($87\pm25~\mu$m) in Martian ripples photographed by the rovers. We then made the double assumption that (i) the grains composing the martian dunes are of the same size and that (ii)  these grains participate to saltation transport whenever the wind is sufficiently strong. In order to support hypothesis (ii), we have computed the transport thresholds in the Martian conditions. We find that the grains which move first for an increasing wind are around $65~\mu$m in diameter. The full discussion of the transport thresholds is however too heavy to be incorporated into the body of the article, mainly devoted to the dune wavelength scaling law. We give below a short but self-sufficient derivation of the scaling laws for the static and dynamic transport thresholds. 

\subsection{Definition of transport thresholds, hysteresis}
We consider the generic case of a fluid boundary layer over a flat bed composed of identical sand grains. For given grains and surrounding fluid, the shear stress $\tau$ controls the sand transport.The dominant mechanism for grain erosion depends on the sand to fluid density ratio. In dense fluids grains are directly entrained by the flow whereas in low density fluids grains are mostly splashed up by other grains impacting the sand bed. Two thresholds are associated to these two mechanisms: starting from a purely static sand bed, the first grain is dragged from the bed and brought into motion at the static threshold $\tau_{\rm sta}$. Once sand transport is established, it can sustain by the collision/ejection processes down to a second threshold $\tau_{\rm dyn}$. The sand transport thus presents an hysteresis -- responsible for instance for the formation of streamers. Finally, when the fluid velocity becomes sufficiently high, more and more grains remain in suspension in the flow, trapped in large velocity fluctuations. Although there is no precise threshold associated to suspension, one expect that these fluctuations becomes dominant for the grain trajectories when the shear velocity $u_*=\sqrt{\tau/\rho_f}$ is much larger than the grain sedimentation velocity $u_{\rm fall}$. The behaviour of these thresholds in the Martian atmosphere, as well as in water and air, with respect to the grain diameter is summarized in figures \ref{ShieldsMars} and \ref{ShieldsWaterAir}, and we shall now discuss how to compute these curves.

\subsection{An analytic expression for the turbulent boundary layer velocity profile}
In the dune context, the flow is generically turbulent far from the sand bed and the velocity profile is known to be logarithmic. However, our problem is more complicated as we have to relate the velocity $u$ of the flow around the sand grains (i.e. the velocity at the altitude, say, $z=d/2$ that we call $v$ in the following) to the shear velocity $u_*$. Depending on the grain based Reynolds number, there either exists a viscous sub-layer between the soil and the fully turbulent zone, or the momentum transfer is directly due to the fluctuations induced by the soil roughness. The first step is thus to derive an expression for the turbulent boundary layer wind profile valid in the two regimes, far or close to the bed. To do so, we express the shear stress as the sum of a visous and a turbulent part as
\begin{equation}
\tau = \rho_f \nu \partial_z u +\kappa^2 (z+r d)^2\rho_f  |\partial_z u | \partial_z u,
\label{tauV+T}
\end{equation}
where $\nu$ is the kinematic viscosity of the fluid, $\kappa$ the von K\'arm\'an constant and $r$ the aerodynamic roughness rescaled by the grain diameter $d$. The shear stress is constant all through the turbulent boundary layer and equal to $\rho_f u_*^2$. Let us define the grain Reynolds number as:
\begin{equation}
Re_0=\frac{2\kappa u_* r d}{\nu} \, .
\end{equation}
and a non-dimensional distance to the bed:
\begin{equation}
Z=1+\frac{z}{r d}
\end{equation}
Note that $Z$ is tends to $1$ on the sand bed and that it is equal to $1+1/2r$ at the center of the grain. With this notations, the differential equation (\ref{tauV+T}) can be rewritten under the form:
\begin{equation}
Z^2 \left(\frac{d (\kappa u/u_*)}{dZ}\right)^2 + \frac{2}{Re_0} \left(\frac{d (\kappa u/u_*)}{dZ}\right) -1 = 0,
\end{equation}
which is easily integrated into:
\begin{equation}
u=\frac{u_*}{\kappa}
\left[
\sinh ^{-1}(Re_0 \bar{Z})+\frac{1-\sqrt{1+Re_0^2 \bar{Z}^2}}{Re_0 \bar{Z}}
\right]_{\bar{Z}=1}^{\bar{Z}=Z}.
\label{windprofile}
\end{equation}
Performing expansions in the purely viscous and turbulent regimes, one can show that a good approximation of the relation between the shear velocity $u_*$ and typical velocity around the grain $v$ is given by
\begin{equation}
u_*^2=\frac{2 \nu }{d}  v+ \frac{\kappa^2}{\ln^2(1+1/2r)} v^2.
\end{equation}

\subsection{Static threshold: influence of Reynolds number}

\begin{figure}[t!]
\begin{center}
\epsfig{file=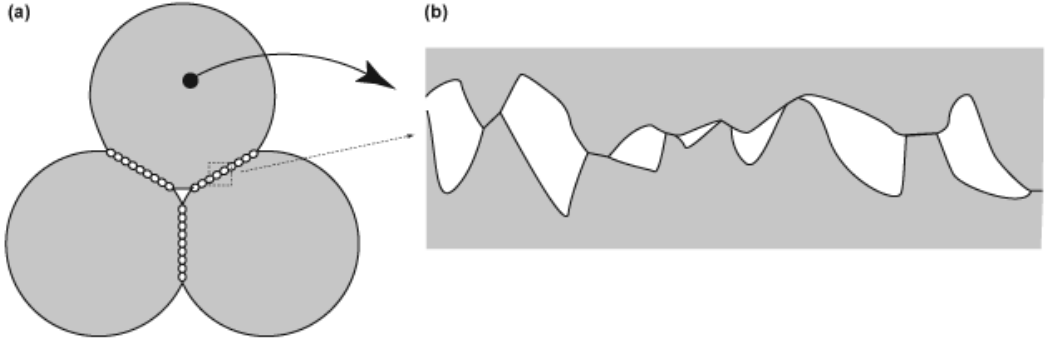,width=\linewidth}
\end{center}
\caption{{\bf a} Grain packing geometry considered for the computation of the static transport thresholds. {\bf b} Schematic drawing showing the contact between grains at the micron scale.}
\label{Threshold}
\end{figure}

The bed load (tractation) threshold is directly related to the fact that the grains are trapped in the potential holes created by the grains at the sand bed surface. To get the scaling laws, the simplest geometry to consider is a single spherical grain jammed between the two neighbouring (fixed) grains below it, see figure~\ref{Threshold}a. Let us first discuss the situation in which the cohesive forces between the grains are negligible and the friction at the contacts is sufficient to prevent sliding. It can be inferred from figure~\ref{Threshold}a that the loss of equilibrium occurs for a value of the driving force $F$ proportional to the submerged weight of the grain: $F \propto ( \rho_s - \rho_f )g d^3$, where $\rho_s$ is the mass density of the sand grain, and $\rho_f$ is that of the fluid -- which is negligible with respect to $\rho_s$ in the case of aeolian transport. As $F$ is proportional to the shear force $\tau d^2$ exerted by the fluid, the non-dimensional parameter controlling the onset of motion is the Shields number, which characterises the ratio of the driving shear stress to the normal stress:
\begin{equation}
\Theta = \frac{\tau}{(\rho_s-\rho_f)gd}.
\end{equation}
The threshold value can be estimated from the geometry of the piling, and depends on whether rolling or lifting is the mechanism which makes the grain move. Finally, the local slope of the bed modifies the threshold value as traps between the grains are less deep when the bed is inclined. In particular, its value must vanish as the bed slope approaches the (tangent of the) avalanche angle. We here ignore these refinements which can be incorporated into the values of $A$ and $B$ (see section \ref{instab}).

At the threshold, the horizontal force balance on a grain of the bed reads
\begin{equation}
\frac{\pi}{6}  \mu ( \rho_s - \rho_f ) g d^3= \frac{\pi}{8}  C_d  \rho_f v_{\rm th}^2 d^2,
\end{equation}
where $\mu$ is a friction coefficient, and $C_d$ the drag coefficient which is a function of the grain Reynolds number. With a good accuracy, the drag law for natural grains can be put under the form:
\begin{equation}
C_d = \left(C_\infty^{1/2}+s \sqrt{\frac{\nu}{v d}}\right)^2
\label{dragcoeff}
\end{equation}
with $C_\infty \simeq 1$ and $s \simeq 5$ for natural sand grains \cite{FC04}.

At this stage, we introduce the viscous size $d_\nu$, defined as the diameter of grains whose
free fall Reynolds number $u_{\rm fall}d/\nu$ is unity:
\begin{equation}
d_\nu=(\rho_s/\rho_f-1)^{-1/3}~\nu^{2/3}~g^{-1/3}
\end{equation}
It corresponds to a grain size at which viscous and gravity effects are of the same order of magnitude. We define $\tilde v$ as the fluid velocity at the scale of the grain normalised by $(\rho_s/\rho_f-1)^{1/2}(gd)^{1/2}$. From the three previous relations, we get the equation for ${\tilde v}_{\rm th}$:
\begin{equation}
C_\infty^{1/2}{\tilde v}_{\rm th} +s  \left(\frac{d_\nu }{d}\right)^{3/4}  {\tilde v}_{\rm th}^{1/2}-\left(\frac{4\mu}{3}\right)^{1/2}=0,
\end{equation}
which solves into
\begin{equation}
{\tilde v}_{\rm th}=\frac{1}{4C_\infty}\left[\left( s^2\left(\frac{d_\nu }{d}\right)^{3/2} +8 \left(\frac{\mu C_\infty}{3}\right)^{1/2}\right)^{1/2}-s  \left(\frac{d_\nu }{d}\right)^{3/4}\right]^2.
\end{equation}
The expression of the static threshold Shields number is finally:
\begin{equation}
\Theta_{\rm th}^{\infty} =2 \left(\frac{d_\nu }{d}\right)^{3/2}  {\tilde v}_{\rm th}+  \frac{\kappa^2}{\ln^2(1+1/2r)}  {\tilde v}_{\rm th}^2.
\end{equation}

\subsection{Static threshold: influence of cohesion}
For small grains, the cohesion of the material strongly increases the static threshold shear stress. Evaluating the adhesion force between grains is a difficult problem in itself. We consider here two grains at the limit of separation and we assume that the multi-contact surface between the grains has been created with a maximum normal load $N_{\rm max}$, see figure~\ref{Threshold}b. The adhesion force $N_{\rm adh}$ can be expressed as an effective surface tension $\tilde{\gamma}$ times the radius of curvature of the contact, which is assumed to scale on the grain diameter:
\begin{equation}
N_{\rm adh} \propto \tilde{\gamma} d.
\end{equation}
This effective surface tension is much smaller than the actual one, $\gamma$, as the real area of contact $A_{\rm real}$ is much smaller than the apparent one  $A_{\rm Hertz}$.:
\begin{equation}
\tilde{\gamma}=\gamma \, \frac{A_{\rm real}}{A_{\rm Hertz}} \, .
\end{equation}
The apparent area of contact can be computed following Hertz law for two spheres in contact under a load $N_{\rm max}$:
\begin{equation}
A_{\rm Hertz} \propto \left(\frac{N_{\rm max} d}{E}\right)^{2/3},
\end{equation}
where E is the Young modulus of the grain \cite{J85}. To express the real area of contact, we need to know whether the micro-contacts are in an elastic or a plastic regime. Within a good approximation, $A_{\rm real}$ can expressed in both cases \cite{G00} as:
\begin{equation}
A_{\rm real} \propto \frac{N_{\rm max}}{M},
\end{equation}
where $M$ is the Young modulus $E$ (elastic regime) or the hardness $H$ (plastic regime) of the material. Altogether, we then get:
\begin{equation}
N_{\rm adh} \propto \gamma \, \frac{E^{2/3}}{M} \, (N_{\rm max} d) ^{1/3}.
\end{equation}

In order to bring into motion such grains, the shear must be large enough to overcome both weight and adhesion, so that, for $N_{\rm max} \sim \rho_s g d^3$ (i.e. the weight of one grain) the critical Shields number is the sum of two terms and takes the form of
\begin{equation}
\Theta_{\rm th} = \Theta_{\rm th}^{\infty} \left [1 + \frac{3}{2} \left ( \frac{d_m}{d} \right )^{5/3} \right ],
\end{equation}
with
\begin{equation}
d_m \propto \left ( \frac{\gamma}{M} \right )^{3/5} \left ( \frac{E}{\rho_s g} \right )^{2/5}.
\end{equation}
Note that, by contrast to the references \cite{IPGW76,PHGI76,IW82,SL00,CGH04}, we find that the adhesive force finally scales, for $d \to 0$ with $d^{4/3}$ and the critical Shields number with $d^{-5/3}$ (instead of exponents $1$ and $-2$ respectively).

\subsection{Dynamical threshold}
The dynamical threshold can be defined as the value of the control parameter for which the saturated flux vanishes. Fitting the flux \emph{vs} shear velocity relation, one can measure the dynamical threshold in a very precise way (see for instance the analysis in \cite{A04} of the data obtained by Rasmussen \emph{et al.} \cite{RIR96}).

\begin{figure}[t]
\begin{center}
\epsfig{file=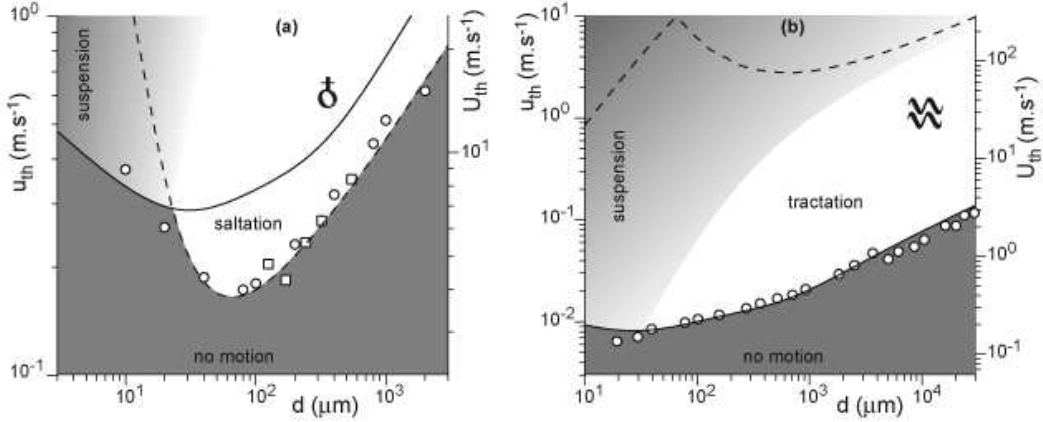,width=\linewidth}
\end{center}
\caption{Diagram showing the mode of transport in the aeolian {\bf (a)} and  underwater {(\bf b)} cases, as a function of the grain diameter $d$ and of the turbulent shear velocity $u_{\rm th}$ (left) or of the wind speed $U_{\rm th}$ at $2$~m above the soil (right). The dynamical threshold (dashed line) is below the static threshold (solid line) in the aeolian case but much above it underwater. The dark gray is the zone where no transport is possible. Above, the background color codes for the ratio $u_*/u_{\rm fall}$: white corresponds to  negligible fluctuations and gray to suspension. The experimental points are taken from ({\Large$\circ$}) Chepil \cite{C45} and ($\square$) Rasmussen \cite{RIR96,A04} in the aeolian case and from  ({\Large$\circ$})  Yalin and Karahan \cite{YK79} in the underwater case.}
\label{ShieldsWaterAir}
\end{figure}

As shown in \cite{A04}, the wind velocity profile is almost undisturbed at the dynamical threshold. The shear stress threshold is thus achieved when the velocity that a grain acquires after a single jump is just sufficient to eject on average one single grain out of the bed after a collision (unit replacement capacity criterion). The impact velocity at this dynamical threshold is thus proportional to the trapping velocity i.e. the velocity needed for a grain to escape from its potential trapping (see Quartier \emph{et al.} \cite{QADD00}):
\begin{equation}
v_\downarrow = a \sqrt{gd \left [1 + \frac{3}{2} \left ( \frac{d_m}{d} \right )^{5/3} \right ]},
\end{equation}
An analytical expression of the dynamical threshold has been derived in  \cite{A04}, but only in the limit of large Reynolds numbers. To generate the plots displayed in figures \ref{ShieldsMars} and \ref{ShieldsWaterAir}, we use here a numerical integration of the equations of motion of a grain -- with equations (\ref{windprofile}) and (\ref{dragcoeff}) for the wind profile and drag coefficient.  The trajectories are computed for an ejection velocity $v_\uparrow$ equal to $e\,v_\downarrow$ and a typical ejection angle $\pi/4$ (see \cite{A04}) .

%

\subsection{From Earth to Mars}
In order to get a reasonable transport diagram on Mars, we have first tuned the different parameters controlling the dynamic and static thresholds to reproduce {\it within the same model} the underwater and aeolian data (figure~\ref{ShieldsWaterAir}). For the friction coefficient $\mu$, we have simply taken the avalanche slope for typical aeolian grains $\tan(32^\circ)$. As already found in \cite{A04}, the rescaled soil roughness $r$ is higher than the value found by Bagnold. It is here adjusted to $r=1/10$.  The value of $a$, the impact velocity needed to eject statistically one grain rescaled by the trapping velocity, is adjusted to $15$. The restitution coefficient was adjusted to $e=2/3$. For obvious reasons it is possible to obtain almost the same curves with different pairs ($a$, $e$). Finally the cohesion length $d_m$ was tuned to $25~\mu$m, which is consistent with the above calculation.

We see in figure~\ref{ShieldsWaterAir} that the agreement is excellent both in the aeolian and underwater cases. Due to the low density ratio in water, the collision process is completely inefficient. As a consequence, the dynamical threshold is well above the static one: the only erosion mechanism  is the direct entrainment of grains by the fluid (tractation). On the other hand, the static threshold --~computed with the same formula as that obtained in water~-- is well above the experimental data in the aeolian case: there is a very important hysteresis in that case. With both sets of data, we have thus a complete calibration of the dynamic and static threshold parameters, which allows to a good degree of confidence for the computation of the transport diagram on Mars (figure~\ref{ShieldsMars}).

\newpage

\end{document}